\documentclass[prc,twocolumn,twoside,showpacs,floatfix]{revtex4}

\usepackage{graphicx,color,rotating}
\usepackage{amsmath}
\usepackage{amssymb,bm}
\usepackage{dcolumn}

\usepackage{times,txfonts}

\newcommand{\eq}[1]{\begin{equation}#1\end{equation}}
\newcommand{\eqmulti}[1]{\begin{equation}\begin{split}#1\end{split}\end{equation}}
\newcommand{\eqalign}[1]{\begin{align}#1\end{align}}

\newcommand{\bra}[1]{\ensuremath{\langle{#1}|\,}}

\newcommand{\ket}[1]{\ensuremath{\,|{#1}\rangle}}

\newcommand{\braket}[2]{\ensuremath{\langle{#1}|{#2}\rangle}}

\newcommand{\matrixe}[3]{\ensuremath{\langle{#1}|\,{#2}\,|{#3}\rangle}}

\newcommand{\op}[1]{\ensuremath{#1}}

\newcommand{\elem}[2]{\ensuremath{{}^{#2}\text{#1}}}

\newcommand{\symboldiamond}[1][black]{{\color{#1}\hspace{-1pt}\scriptsize\begin{turn}{45} $\blacksquare$ \end{turn}}}
\newcommand{\symboltriangle}[1][black]{{\color{#1}$\blacktriangle$}}

\newcommand{\symbolcircle}[1][black]{{\color{#1}$\bullet$}}

\definecolor{FGViolet}{rgb}{0.61,0.32,0.61}
\definecolor{FGDarkBlue}{rgb}{0,0,0.6}
\definecolor{FGBlue}{rgb}{0,0,0.8}
\definecolor{FGLightBlue}{rgb}{0.2, 0.6, 0.8}
\definecolor{FGGreen}{rgb}{0.2,0.7,0.2}
\definecolor{FGLightGreen}{rgb}{0.4,1,0.4}
\definecolor{FGYellow}{rgb}{1,0.95,0}
\definecolor{FGOrange}{rgb}{0.95,0.5,0.1}
\definecolor{FGRed}{rgb}{0.8,0,0}
\definecolor{FGWhite}{rgb}{1,1,1}
\definecolor{FGLightGray}{rgb}{0.8,0.8,0.8}
\definecolor{FGGray}{rgb}{0.5,0.5,0.5}
\definecolor{FGDarkGray}{rgb}{0.3,0.3,0.3}
\definecolor{FGBlack}{rgb}{0,0,0}

\newcommand{\linethindashed}[1][black]{\unitlength 1ex
  {\color{#1}
  \begin{picture}(6,1)
  \linethickness{0.12mm}
  \put(0,0.5){\line(1,0){1.5}}
  \put(2.2,0.5){\line(1,0){1.5}}
  \put(4.4,0.5){\line(1,0){1.5}}
  \end{picture}}\nolinebreak
}

\begin{document}

\title{Spectra of Open-Shell Nuclei with Pad\'e-Resummed Degenerate Perturbation Theory}

\author{Joachim Langhammer}
\email{joachim.langhammer@physik.tu-darmstadt.de}
\author{Robert Roth}
\author{Christina Stumpf}

\affiliation{Institut f\"ur Kernphysik, Technische Universit\"at Darmstadt,
64289 Darmstadt, Germany}

\date{\today}

\begin{abstract}

We apply degenerate many-body perturbation theory at high orders for the ab-initio description of ground states and excitation spectra of open-shell nuclei using soft realistic nucleon-nucleon interactions. We derive a recursive formulation of standard degenerate many-body perturbation theory that enables us to evaluate order-by-order perturbative energy and state corrections up to the 30th order. We study \elem{Li}{6,7} as test cases using a similarity renormalization group (SRG) evolved nucleon-nucleon interaction from chiral effective field theory. The simple perturbation series exhibits a strong, often oscillatory divergence, as was observed previously for ground states of closed-shell nuclei. Even for very soft interactions resulting from SRG evolutions up to large flow parameter, i.e. low momentum scales, the perturbation series still diverges. However, a resummation of the perturbation series via Pad\'e approximants yields very stable and converged ground and excited-state energies in very good agreement with exact no-core shell-model calculations for the same model space.

\end{abstract}


\maketitle


\section{Introduction}

The precise description of the nuclear spectroscopy is one of the major challenges in nuclear structure theory. A variety of many-body approaches, both, exact and approximate, have been developed to tackle this challenge. A simple and well-known tool to obtain approximate solutions of the many-body problem is Rayleigh-Schr\"odinger many-body perturbation theory (MBPT) \cite{Schr26}. Low-order MBPT has been used for studies of systematics of ground-state properties of closed-shell nuclei throughout the nuclear chart \cite{GuRo10, RoNe10, RoPa06}, as well as for infinite neutron and nuclear matter calculations \cite{HeBo11, TeKr12}. Recently, we have used high-order MBPT, i.e. an order-by-order evaluation of the perturbation series up to typically 30th order, to address ground-state energies of light closed-shell nuclei \cite{RoLa10}. It turns out that the perturbation series in general does not converge. However, through a resummation by Pad\'e approximants we have shown that one can utilize the information of the divergent power series to determine the ground-state energy of closed-shell nuclei with the same precision as in the no-core shell-model (NCSM) approach \cite{NaQu09, MaVa12, RoLa11} using the same model space.

Unfortunately, standard nondegenerate Rayleigh-Schr\"odinger perturbation theory is applicable only to ground states of closed-shell nuclei. Both, the step to excited states and the step to open-shell nuclei leads to degeneracies in the unperturbed (zeroth-order) energy level, which must be considered explicitly. In this paper, we investigate degenerate Rayleigh-Schr\"odinger perturbation theory to overcome this limitation. We derive recursive formulas for an efficient order-by-order construction of the perturbation series. With this extension we are able to study ground and excited states of closed and open-shell nuclei.
We focus on the application of degenerate many-body perturbation theory (DMBPT) up to high orders and Pad\'e resummations for the description of ground states and excitation spectra of light nuclei, specifically \elem{Li}{6} and \elem{Li}{7}, and we compare the DMBPT results with exact NCSM calculations for the same model space and Hamiltonian. We employ a nucleon-nucleon interaction from chiral effective field theory \cite{MaEn11,EpHa09} at next-to-next-to-next-to-leading-order (N$^{3}$LO) in the version of Entem and Machleidt \cite{EnMa03} after an additional similarity renormalization group (SRG) transformation \cite{GlWi93, BoFu07, RoNe10} to further soften the interaction. As expected \cite{KvJa11, RoLa10}, the DMBPT series does not converge, not even for very soft interactions. However, we will show that a resummation through Pad\'e approximants leads to stable and accurate predictions for the spectra.

This paper is organized as follows:  In Sec.~\ref{sec:DMBPT}, we present the formalism of DMBPT, derive the relevant formulas for the energy and state corrections, and highlight their recursive structure. The latter allows us to construct the perturbation series up to high orders. In Sec.~\ref{sec:Pade} we introduce the Pad\'e approximants that we use to resum the typically divergent power series from DMBPT. Finally, we show results for the spectra of \elem{Li}{6} and \elem{Li}{7} obtained from the DMBPT power series in Sec.~\ref{sec:results-Li6-7}.


\section{Degenerate Many-Body Perturbation Theory}\label{sec:DMBPT}

The starting point for the formulation of MBPT is the eigenvalue problem of the intrinsic Hamiltonian
\eq{\label{eq:ev-problem}
\op{H} \ket{\Psi_{n}}=\big(\op{T} - \op{T}_{\text{cm}} + \op{V}\big) \ket{\Psi_{n}} = E_{n} \ket{\Psi_{n}}\,,
}
with $T_{\text{cm}}$ denoting the center-of-mass kinetic energy and $\op{T}-\op{T}_{\text{cm}}$ being the intrinsic kinetic energy. Note that the interaction $\op{V}$ is general and might include three-body forces as well. Next, we partition the Hamiltonian $\op{H}$ into an unperturbed part $\op{H}_{0}$ and a perturbation $\op{W}$ with an auxiliary parameter $\lambda$ such that the original Hamiltonian is recovered for $\lambda=1$
\eqmulti{\label{eq:H_lambda}
\op{H}
\;\;\xrightarrow{\text{partitioning}}\;\; 
\op{H}_{\lambda} 
&= \op{H}_{0} + \lambda \op{W} \\
&= \op{H}_{0} + \lambda \big(\op{H} -\op{H}_{0}\big)\,.
}
Formally one has complete freedom in defining this partitioning and, thus, in choosing the unperturbed Hamiltonian $\op{H}_{0}$. In practical applications the choice of $\op{H}_{0}$ is often motivated by computational simplicity. The eigenvalue problem of the unperturbed Hamiltonian 
\eq{\label{eq:unperturbed_evprob}
\op{H}_{0} \ket{\Phi_{n}} = \epsilon_{n} \ket{\Phi_{n}}
}
defines the unperturbed basis $\{\ket{\Phi_{n}}\}$ that is used throughout the perturbative expansion. This basis should be sufficiently easy to handle formally and computationally, but also adequate for the physical system under consideration. For nuclear-structure applications typical choices for $\op{H}_0$ are Hartree-Fock or harmonic-oscillator (HO) single-particle Hamiltonians. Throughout this paper we choose the latter, i.e., the unperturbed states $\{\ket{\Phi_{n}}\}$ are given by Slater determinants of single-particle HO states. The energy eigenvalues $\epsilon_{n}$ are determined by the sum of the single-particle energies of occupied states.  

In the simplest form of MBPT the unperturbed state that represents the eigenstate of interest is required to be nondegenerate. When using the HO basis, this is true only for the ground states of light doubly-magic nuclei, e.g. \elem{He}{4}, \elem{O}{16} and \elem{Ca}{40} \cite{RoLa10}. However, for open-shell nuclei or excited states of closed-shell nuclei the unperturbed states exhibit degeneracies and one has to resort to degenerate Rayleigh-Schr\"odinger many-body perturbation theory. In the following, we derive an iterative formulation of DMBPT that enables us to efficiently evaluate perturbative corrections to the energies and states up to very high orders.

To characterize the degeneracy of the unperturbed states, we introduce an additional degeneracy index $d$ for the unperturbed  Slater determinants $\ket{\Phi_{nd}}$, which labels the states spanning the $g_n$-dimensional degenerate subspace associated with the energy eigenvalue $\epsilon_{n}$. As a consequence, the index $d$ also appears in the power-series for the perturbed energies and states
\eqalign{\label{eq:power_series_enery}
E_{nd}(\lambda)    &= \epsilon_{n} + \lambda E_{nd}^{(1)} + \lambda^{2} E_{nd}^{(2)} + \ldots \\\label{eq:power_series_state}
\ket{\Psi_{nd}(\lambda)} &= \ket{\Psi_{nd}^{(0)}} + \lambda \ket{\Psi_{nd}^{(1)}} + \lambda^{2} \ket{\Psi_{nd}^{(2)}} + \ldots \,,
}
with $E_{nd}^{(0)} = \epsilon_n$.
We insert this ansatz in the eigenvalue problem of the Hamiltonian \eqref{eq:H_lambda} and match same orders of $\lambda$. For order $\lambda^{0}$ we recover the unperturbed eigenvalue problem \eqref{eq:unperturbed_evprob}, whereas for order $\lambda^{p}$ with $p\geq1$ we obtain
\eq{\label{eq:maineq}
\op{W} \ket{\Psi_{nd}^{(p-1)}} + \op{H}_{0} \ket{\Psi_{nd}^{(p)}} = \sum_{j=0}^{p} E_{nd}^{(j)} \ket{\Psi_{nd}^{(p-j)}} \,.
}
The unperturbed states $\ket{\Psi_{nd}^{(0)}}$ obviously enter in Eq.~\eqref{eq:maineq}. However, for each $g_n$-dimensional degenerate subspace for an unperturbed energy $\epsilon_{n}$ we can choose arbitrary linear combinations of the naive Slater determinants $\ket{\Phi_{nd}}$ to represent the unperturbed states 
\eq{\label{eq:linear_combination_degenerate}
\ket{\Psi_{nd}^{(0)}} = \sum_{e=0}^{g_{n}-1} \chi_{nd,ne} \ket{\Phi_{ne}} \,,
}
In order to fix the expansion coefficients $\chi_{nd,ne}$ we consider Eq.~\eqref{eq:maineq} for $p=1$, insert the expansion  \eqref{eq:linear_combination_degenerate} and multiply with $\bra{\Phi_{n d'}}$, yielding
\eq{\label{eq:ev_degenerate}
\sum_{e=0}^{g_{n}-1} \big(\, \matrixe{\Phi_{n d'}}{W}{\Phi_{n e}} - E_{n d}^{(1)} \delta_{d' e}\,\big)\; \chi_{nd,ne} =0\,,
}
where we used the orthogonality $\braket{\Phi_{n d}}{\Phi_{n d'}} = \delta_{d d'}$.
Equation~\eqref{eq:ev_degenerate} is an eigenvalue equation in the degenerate subspace for $\epsilon_n$. The eigenvectors define the expansion coefficients $\chi_{nd,ne}$ of the unperturbed states in Eq.~\eqref{eq:linear_combination_degenerate}, and the eigenvalues the first-order energy corrections $E_{nd}^{(1)}$. Moreover, the following relations hold for the unperturbed states $\ket{\Psi_{nd}^{(0)}}$ 
\eq{
\braket{\Psi_{nd}^{(0)}}{\Psi_{nd'}^{(0)}} 
= \sum_{e=0}^{g_{n}-1} \chi_{nd,ne}^{*}\; \chi_{nd',ne} = \delta_{dd'}\,.
}
Using the intermediate normalization $\braket{\Psi_{nd}^{(0)}}{\Psi_{nd}} = 1$ we obtain from Eq.~\eqref{eq:power_series_state} the relation
\eq{
\braket{\Psi_{nd}^{(0)}}{\Psi_{nd}^{(p)}} = \delta_{0p} \;.
}
After multiplying Eq.~\eqref{eq:maineq} with $\bra{\Psi^{(0)}_{nd}}$ and using the previous orthogonality relation, we obtain a simple expression for the $p$-th order energy correction
\eq{\label{eq:pth-order_energy}
E_{nd}^{(p)} = \matrixe{\Psi^{(0)}_{nd}}{\op{W}}{\Psi_{nd}^{(p-1)}}\,,
}
which has the same form as in nondegenerate MBPT \cite{RoLa10}. The derivation of the perturbative corrections to the states $\ket{\Psi_{nd}^{(p)}}$ is more involved. We start off with formally expanding the $p$-th order state corrections for $p\geq 1$ in terms of the unperturbed basis. Within the degenerate subspace $n$ of the target state we use the unperturbed basis $\ket{\Psi_{ne}^{(0)}}$ of Eq.~\eqref{eq:linear_combination_degenerate}, for the orthogonal subspaces $m\neq n$ we use the naive unperturbed basis of Slater determinants $\ket{\Phi_{me}}$ for simplicity. Thus we obtain the following expansion of the $p$-th order perturbative correction
\eq{\label{eq:exp_unperturbed}
\ket{\Psi_{nd}^{(p)}} 
= \sum_{m}^{m\neq n} \sum_{e} \ket{\Phi_{me}}\braket{\Phi_{me}}{\Psi_{nd}^{(p)}} 
+ \sum_{e}^{e\neq d} \ket{\Psi^{(0)}_{ne}}\braket{\Psi^{(0)}_{ne}}{\Psi_{nd}^{(p)}}\,,
}
The remaining task is to derive expressions for the expansion coefficients
\eq{
C_{nd,me}^{(p)}=\braket{\Phi_{me}}{\Psi_{nd}^{(p)}} \quad m\neq n\,,
}
and
\eq{
D_{nd,ne}^{(p)}=\braket{\Psi^{(0)}_{ne}}{\Psi_{nd}^{(p)}} \quad e\neq d \,.
}
We multiply Eq.~\eqref{eq:maineq} with $\bra{\Phi_{me}}$ for $m\neq n$, use Eq.~\eqref{eq:unperturbed_evprob} and obtain
\eq{\label{eq:C-coeff}
C_{nd,me}^{(p)} = \frac{1}{\epsilon_{n} - \epsilon_{m}} \left( \matrixe{\Phi_{me}}{\op{W}}{\Psi_{nd}^{(p-1)}} - \sum_{j=1}^{p-1}E_{nd}^{(j)} C_{nd,me}^{(p-j)} \right)\,.
}
To get the coefficients $D_{nd,ne}^{(p)}$ we multiply Eq.~\eqref{eq:maineq} with $\bra{\Psi_{ne}^{(0)}}$ for $e\neq d$ and make use of the expansion~\eqref{eq:exp_unperturbed} and our results from the diagonalization in the degenerate subspace, yielding
\eqmulti{\label{eq:D-coeff}
D_{nd,ne}^{(p)} = \frac{1}{E_{nd}^{(1)} - E_{ne}^{(1)}} \left(\sum_{\begin{subarray}{c} m,e'\end{subarray}}^{m\neq n} \matrixe{\Psi^{(0)}_{ne}}{\op{W}}{\Phi_{me'}} C_{nd,me'}^{(p)} \right .  \\  \left . - \sum_{j=1}^{p-1} E_{nd}^{(j+1)} D_{nd, ne}^{(p-j)}\right)\,.
}
Note that Eqs.~\eqref{eq:C-coeff} and \eqref{eq:D-coeff} hold for $p\geq 1$.

For implementation we manipulate Eqs.~\eqref{eq:pth-order_energy}, \eqref{eq:C-coeff}, and \eqref{eq:D-coeff} into a more convenient form, using matrix elements of $\op{W}$ with respect to the naive Slater determinants $\ket{\Phi_{nd}}$ as input. 
We obtain for the $p$th-order perturbative energy correction
\eq{\label{eq:recursive_energy_correction}
E_{nd}^{(p=1)}= \sum_{e=0}^{g_n-1} \sum_{e'=0}^{g_n-1} \chi_{nd,ne}^{\ast}\,\chi_{nd,ne'} \matrixe{\Phi_{ne}}{\op{W}}{\Phi_{ne'}}\,,}
and 
\eq{\label{eq:recursive_energy_correctionp2}
\displaystyle E_{nd}^{(p\geq2)}=\sum_{\begin{subarray}{c}m,e \end{subarray}}^{m\neq n} \sum_{e'=0}^{g_n-1} \chi_{nd,ne'}^{\ast} \matrixe{\Phi_{ne'}}{\op{W}}{\Phi_{me}} \cdot C_{nd,me}^{(p-1)} \,. 
}
The expressions for the $C$-coefficients are recast to $C_{nd,me}^{(p=0)} = 0$,
\eq{
C_{nd,me}^{(p=1)} =
  \dfrac{\matrixe{\Phi_{me}}{\op{W}}{\Phi_{nd}}}{\epsilon_{n}-\epsilon_{m}}
  \,,}
and
\begin{align} \label{eq:Ccoeffp2}
   C_{nd,me}^{(p\geq2)} = &\dfrac{1}{\epsilon_{n} - \epsilon_{m}} \cdot \left( \displaystyle\sum_{\begin{subarray}{c}m^{\prime}, e^{\prime} \end{subarray}}^{m^{\prime} \neq n } \matrixe{\Phi_{me}}{\op{W}}{\Phi_{m^{\prime}e^{\prime}}} C_{nd,m^{\prime}e^{\prime}}^{(p-1)} \right . \notag \\
    &\left . +  \displaystyle\sum_{\begin{subarray}{c}e^{\prime}\end{subarray}}^{e^{\prime}\neq d}  \sum_{e'' = 0}^{g_n-1} \chi_{ne',ne''} \matrixe{\Phi_{me}}{\op{W}}{\Phi_{ne^{\prime\prime}}} D_{nd,ne^{\prime}}^{(p-1)}\right. \notag\\ &\left.- \displaystyle\sum_{j=1}^{p-1} E_{nd}^{(j)} C_{nd,me}^{(p-j)} \right ) \,.
\end{align}
For the $D$-coefficients we get $D_{nd,ne}^{(p=0)} = 0$ and
\begin{align}\notag
\label{eq:D-coeff-final}
D_{nd,ne}^{(p \geq1)} = 
\dfrac{1}{E_{nd}^{(1)} - E_{ne}^{(1)}} \cdot &\left( \displaystyle \sum_{\begin{subarray}{c} m, e^{\prime}\end{subarray}}^{ m\neq n} \sum_{e''=0}^{g_n-1} \chi_{ne,ne''}^{\ast} \matrixe{\Phi_{ne''}}{\op{W}}{\Phi_{me^{\prime}}} C_{nd,me^{\prime}}^{(p)}\right.\\\hspace*{-20pt}&\left. - \displaystyle\sum_{j=1}^{p-1} E_{nd}^{(j+1)} D_{nd,ne}^{(p-j)} \right )\,.
\end{align}

For the construction of the perturbation series up to high orders, carrying out all summations explicitly is very inefficient. Closer inspection of Eqs.~(\ref{eq:recursive_energy_correction}) -- (\ref{eq:D-coeff-final}) reveals their recursive structure: For the $p$th-order energy correction~\eqref{eq:pth-order_energy} we need the state correction of $(p-1)$th-order. However, we see from Eqs.~\eqref{eq:recursive_energy_correction} and \eqref{eq:recursive_energy_correctionp2} that the energy correction depends only the coefficients $C_{nd,me}^{(p-1)}$ which implicitly require the coefficients $D_{nd,ne}^{(p-2)}$. The expansion coefficients $C_{nd,ne}^{(p\geq2)}$ themselves depend on $C$-coefficients, $D$-coefficients and energy corrections of lower order, while the coefficients $D_{nd,ne}^{(p\geq1)}$ depend on lower-order $D$-coefficients and on same-order $C$-coefficients. To proceed to high orders it is indispensable to make use of this recursive structure, i.e. to construct the perturbation series order by order.

We start the construction of the perturbation series for the energy with the zeroth-order contribution simply given by the HO energy of $\ket{\Psi_{nd}^{(0)}}$. We obtain the first correction $E_{nd}^{(1)}$ from the diagonalization in the degenerate subspace. To go to second-order energy correction we need the first-order coefficients $C_{nd}^{(1)}$ only. To compute the energy correction of order $p$ with $p\ge3$ we first calculate the coefficients $D_{nd,ne}^{(p-2)}$ for which we need the already known $C_{nd}^{(p-2)}$ and all previous $D$-coefficients. Then, we use the $D_{nd,ne}^{(p-2)}$ coefficients and all $C$-coefficients up to order $(p-2)$ to calculate the coefficients $C_{nd}^{(p-1)}$. According to Eq.~\eqref{eq:recursive_energy_correctionp2}, the coefficients $C_{nd}^{(p-1)}$ enable the computation of the $p$th-order energy correction $E_{nd}^{(p)}$. This scheme allows for the iterative setup of high-order DMBPT. We present our results  based on DMBPT corrections up to 30th order for excited states of \elem{Li}{6} and \elem{Li}{7} and the according spectra in section~\ref{sec:results-Li6-7}.

From a computational point of view, the critical operations that have to be performed for evaluating the perturbative corrections are matrix-vector multiplications of the Hamilton matrix and the coefficient vectors, see e.g. Eqs.~\eqref{eq:recursive_energy_correctionp2} and \eqref{eq:Ccoeffp2}. The coefficient vectors of all previous orders need to be stored. This is similar to a simple Lanczos algorithm \cite{Lanc50} for determining the few lowest eigenvalues and the corresponding eigenvectors of the many-body Hamiltonian matrix, as used in the NCSM. Therefore, the limitations in terms of particle number and model-space size for high-order DMBPT calculations are similar to those of the NCSM. In order to apply DMBPT at high orders beyond the domain of NCSM-type calculations, one has to devise alternative ways of efficiently carrying out the nested sums involved in the expressions above.


\section{Pad\'e Approximants}\label{sec:Pade}

Even though the recursive formulation of DMBPT will allow us to perform calculations up to very high order of the perturbation series, simple partial summations of this series do not guarantee convergence of the energies. On the contrary, for ground-state energies of closed-shell nuclei we have shown \cite{RoLa10} that even soft interactions lead to strongly diverging power series at the physical point $\lambda=1$. In this sense MBPT does not converge in many practically relevant cases.

However, the coefficients of the power series of MBPT contain all the relevant physics information on the problem. An efficient and elegant way to extract this information are Pad\'e approximants \cite{BaGr96, Bake65}. Instead of expanding the function $E(\lambda)$ in a simple power series, we can employ a slightly more involved ansatz and represent $E(\lambda)$ as a rational function of two power series in $\lambda$
\eq{
E(\lambda) = \dfrac{a_0  + \lambda a_1 + \lambda^2 a_2 \ldots}
                            {b_0  + \lambda b_1  + \lambda^2 b_2 \ldots}\,.
}
If we truncate the power series in the numerator at order $L$ and in the denominator at order $M$ this defines the so-called Pad\'e approximant  
\eq{ \label{eq:pade}
[L/M](\lambda) = \dfrac{a_0  + \lambda a_1 + \lambda^2 a_2 \ldots + \lambda^L a_L}
                            {b_0  + \lambda b_1  + \lambda^2 b_2 \ldots + \lambda^M b_M}\,.
}
In order to determine the unknown coefficients $a_i$ and $b_j$ from the known coefficients $E^{(p)}$ of the simple MBPT power series \eqref{eq:power_series_enery}, we use a Taylor expansion of \eqref{eq:pade} and extract equations connecting the two sets of coefficients by matching the different orders in $\lambda$. Starting from a perturbation series up to order $p$ this allows us to extract the coefficients of Pad\'e approximants with $L+M\leq p$ by solving a set of coupled linear equations. An equivalent but more elegant way to compute the Pad\'e approximant uses the ratio of determinants constructed directly from  the corrections $E^{(p)}$ of the MBPT series \cite{BaGr96}
\eq{ \label{eq:padeapprox}
[L/M](\lambda) = \small
\frac{\left|
\begin{array}{cccc}
E^{(L-M+1)} & E^{(L-M+2)} & \cdots & E^{(L+1)}\\
E^{(L-M+2)} & E^{(L-M+3)} & \cdots & E^{(L+2)}\\
\vdots & \vdots & \ddots & \vdots\\
E^{(L)} & E^{(L+1)} & \cdots & E^{(L+M)}\\
\sum\limits_{p=0}^{L-M} E^{(p)} \lambda^{M+p} & \sum\limits_{p=0}^{L-M+1} E^{(p)}\lambda^{M+p-1} & \cdots & \sum\limits_{p=0}^{L} E^{(p)}\lambda^{p}
\end{array}
\right|}
{\left|
\begin{array}{ccccc}
E^{(L-M+1)} & E^{(L-M+2)} & \cdots & E^{(L+1)}\\
E^{(L-M+2)} & E^{(L-M+3)} & \cdots & E^{(L+2)}\\
\vdots & \vdots & \ddots & \vdots\\
E^{(L)} & E^{(L+1)} & \cdots &  E^{(L+M)}\\
\lambda^{M} & \lambda^{M-1} & \cdots & 1
\end{array}
\right|}\,,\raisetag{15pt}
}
with $E^{(p)}=0$ for $p<0$. Evaluating the above determinants for the physical point $\lambda=1$ directly provides us with a Pad\'e resummed $[L/M]$ approximation for the ground- and excited-state energies using the perturbative corrections $E^{(p)}$ up to order $p=L+M$.

As we have shown in Ref. \cite{RoLa10}, Pad\'e approximants provide a reliable tool to obtain the ground-state energy of doubly-magic nuclei in excellent agreement with exact NCSM calculations even if the simple MBPT power series exhibits a strongly diverging behavior. Using the recursive formulation of DMBPT derived above, we will extend these studies in the following section to ground states and excitation spectra of open-shell systems using the example of \elem{Li}{6} and \elem{Li}{7}.


\section{Spectra of \elem{Li}{6} and \elem{Li}{7}}
\label{sec:results-Li6-7}

\begin{figure*}[t]
\centering\includegraphics[width=1\textwidth]{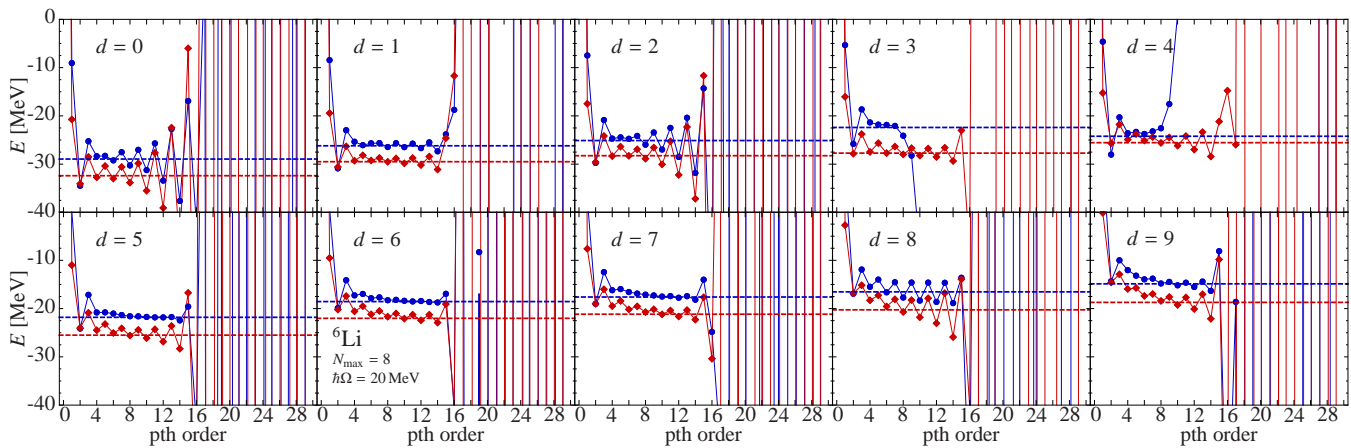}
\caption{(color online) Energy from DMBPT power series truncated at order $p$ for the energy levels of \elem{Li}{6} corresponding to the degenerate HO $n=0$ subspace  using a $N_{\text{max}} = 8$ model space for $\hbar\Omega=20\,\text{MeV}$. The indices $d$ are determined by diagonalization in the degenerate subspace, i.e. by the first-order energy corrections. The dashed lines correspond to the NCSM result in the same model space. Two SRG flow parameters are shown: $0.04\,\text{fm}^{4}$ ($\Lambda\approx2.24\,\text{fm}^{-1}$; \symbolcircle[FGBlue], \linethindashed[FGBlue]) and the softer $0.16\,\text{fm}^{4}$ ($\Lambda\approx1.58\,\text{fm}^{-1}$;\,\symboldiamond[FGRed], \linethindashed[FGRed]).}
\label{fig:DRPT_Li6}
\end{figure*}

As a first application and benchmark of the recursive formulation of DMBPT for open-shell nuclei and excitation spectra, we consider \elem{Li}{6} and \elem{Li}{7} as test cases. In these systems exact NCSM calculations are easily possible and serve as a reference to study the accuracy of the DMBPT results. Therefore, we perform the DMBPT calculations using the same many-body basis and model-space truncation as used in the NCSM. The unperturbed Hamiltonian $H_0$ consists of the kinetic energy and a one-body harmonic oscillator potential leading to an unperturbed basis $\ket{\Phi_{nd}}$ of Slater determinants consisting of harmonic-oscillator single-particle states. As in the NCSM, we truncate the Hilbert space to a finite model space by imposing a maximum excitation energy $N_{\text{max}}\hbar\Omega$ above the lowest unperturbed energy \cite{NaQu09, MaVa12, RoLa11} . In this unperturbed basis, degeneracy emerges if we study ground states of nuclei away from the harmonic-oscillator shell closures or if we investigate excited states.

For the full Hamiltonian $H$ given in Eq.~\eqref{eq:ev-problem} we include a two-nucleon interaction $V$ based on chiral effective field theory at next-to-next-to-next-to leading order (N$^{3}$LO). Starting from the N$^3$LO interaction of Entem and Machleidt \cite{EnMa03} we apply the Similarity Renormalization Group (SRG) \cite{GlWi93, BoFu07, RoNe10} to soften the interaction by a continuous unitary transformation. As a result, the full Hamiltonian has very favorable convergence properties as we increase the size of the many-body model space by increasing $N_{\max}$. Furthermore, the unitary transformation is believed to facilitate the order-by-order convergence of MBPT \cite{BoSc05}, which we will come back to later on. 

First, we investigate \elem{Li}{6} in the framework of DMBPT up to 30th order. The degenerate subspace for the lowest unperturbed energy ($n=0$), which is simply the $N_{\max}=0$ subspace in the language of the NCSM, consists of 10 Slater determinants $\ket{\Phi_{0d}}$. In a first step, we diagonalize the perturbation $W$ in this subspace yielding the first-order energy corrections $E_{0d}^{(1)}$ and the unperturbed basis $\ket{\Psi_{0d}^{(0)}}$. We assign the degeneracy index $d=0,...,9$ in ascending order of the first-order energy. Then we use the recursive formulation of DMBPT derived in Sec. \ref{sec:DMBPT} to compute the perturbative corrections to the energies and states up to 30th order. 
 
The results of the DMBPT calculations for the \elem{Li}{6} energies of all states with $n=0$ are shown in Fig.~\ref{fig:DRPT_Li6} for a model space with $N_{\text{max}}=8$ and $\hbar\Omega=20\,\text{MeV}$. We use two different Hamiltonians including SRG-evolved chiral NN interactions with SRG flow-parameters $\alpha=0.04\,\text{fm}^4$ ($\Lambda\approx2.24\,\text{fm}^{-1}$) and $0.16\,\text{fm}^4$ ($\Lambda\approx1.58\,\text{fm}^{-1}$), respectively. For comparison we show the exact NCSM results for those Hamiltonians in the same model space as horizontal lines. In all cases the perturbation series diverges. However, we can distinguish different characteristics of the partial sums of the perturbation series as function of the truncation order $p$. 

Let us first consider the harder interaction with $\alpha=0.04\,\text{fm}^4$ (blue discs). One class of states ($d=1,5,6,7,9$) exhibits an apparent alternating convergence with increasing order $p$ up to $p\approx12$, but then at $p\approx 16$ the size of the perturbative corrections explodes and the partial sum diverges in an oscillatory pattern. For another class ($d=0,2,8$) the oscillatory behavior sets in earlier and the amplitude first increases slowly before the rapid divergence sets in. A third class ($d=3,4$) diverges monotonously starting already at $p\approx8$. 

The general situation is the same for the second Hamiltonian using a SRG-transformed chiral NN interaction with $\alpha=0.16\,\text{fm}^4$ (red diamonds). This interaction is generally considered to be very soft and shows a rapid convergence of the energies as function of model space size. Sometimes these interactions are termed 'perturbative', based on an analysis of Weinberg eigenvalues in two-body systems \cite{BoFu10,BoSc05}. Our order-by-order calculation in DMBPT up to $p=30$ shows that the softness of the interaction does not guarantee convergence---not even a systematic improvement of the convergence behavior. For all states we observe a strong oscillatory divergence for high orders of DMBPT. Thus, in terms of the order-by-order perturbation theory for a light nucleus, even these very soft interactions lead to a divergent perturbation series and are non-perturbative in this sense. 

The divergence of the perturbation series makes it impossible to determine a robust and unambiguous approximation for the exact eigenvalues from the high-order results. This problem was already found and addressed in our previous study focusing on ground-state energies of closed-shell nuclei \cite{RoLa10}. 

Even low-order estimates obtained from the second- or third-order calculations do not provide a reliable guideline. For the harder interaction with $\alpha=0.04\,\text{fm}^4$ the exact NCSM eigenvalue is typically between the second-order and the third-order estimate, i.e. the second-order approximation gives an energy below and the third-order approximation an energy above the exact energy eigenvalue. However, with increasing degeneracy index $d$ the second-order approximation moves up and appears above the exact eigenvalue for $d=9$. This trend is even more pronounced for the soft interaction with $\alpha=0.16\,\text{fm}^4$, starting from $d=5$ the second-order approximation and all other low-order approximations are above the exact energy. In particular, going from second- to third-order DMBPT enlarges the discrepancy to the exact NCSM result. Thus, in general, low-order approximations do not provide a controlled estimate for the exact eigenvalue. 

\begin{figure*}[t]
\centering\includegraphics[width=1\textwidth]{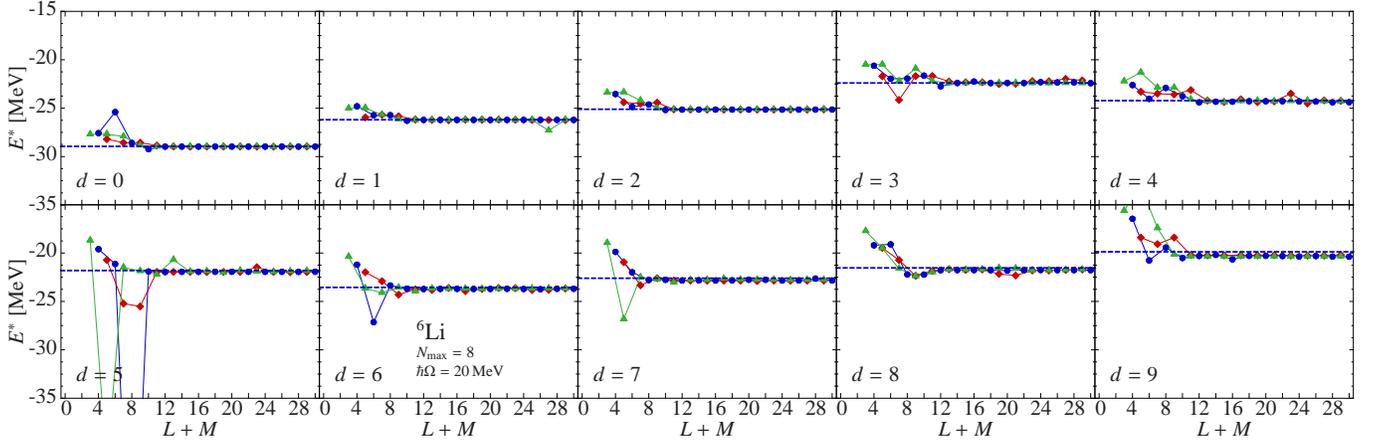}
\caption{(color online) Pad\'e approximation for energy of the ten states corresponding to the $n=0$ HO subspace of \elem{Li}{6}. Shown are the diagonal $[L/L]$ (\symbolcircle[FGBlue]), super-diagonal $[L/L+1]$ (\symboldiamond[FGRed]) and sub-diagonal $[L/L-1]$ (\symboltriangle[FGGreen]) Pad\'e approximants. The dashed horizontal lines represent the exact energies obtained by a NCSM calculation. We used a $N_{\text{max}}=8$ model space with HO frequency $20\,\text{MeV}$ and the SRG parameter is $0.04\,\text{fm}^{4} (\Lambda\approx2.24\,\text{fm}^{-1})$.}
\label{fig:Pade_Li6}
\end{figure*}

To overcome the convergence problems of the simple perturbation series defined in Eq.~\eqref{eq:power_series_enery}, we compute the Pad\'e approximants according to Eq.~\eqref{eq:padeapprox} and evaluate them at $\lambda=1$. The diagonal $[L/L]$, the super-diagonal $[L/L+1]$ and the sub-diagonal $[L/L-1]$ Pad\'e approximants are shown in Fig.~\ref{fig:Pade_Li6} again for the ten states corresponding to the lowest degenerate subspace $n=0$ of \elem{Li}{6} for the interaction with SRG parameter $0.04\,\text{fm}^{4}$. Recall that the information that enters the $[L/M]$ Pad\'e approximant is exactly the same as in the perturbation series truncated at $(L+M)$th order. 

If we use only low-order DMBPT energy corrections as input the quality of the Pad\'e approximants is comparable to the simple perturbation series truncated at these orders. 
If we include information from higher orders of DMBPT, the agreement of the Pad\'e approximants with the exact NCSM results improves successively.
Beyond $L+M=15$ we generally find an excellent agreement of the Pad\'e approximants with the exact NCSM result for the same model space. 
This can be understood in terms of the Pad\'e conjecture, which postulates the existence of a convergent subsequence of diagonal Pad\'e approximants \cite{BaGr96, Bake65, RoLa10}. Nonetheless, there are individual approximants that show larger deviations from the exact result, e.g. for $d=3, 4, 5$ or $8$. However, we observe deviations only for non-diagonal approximants, which are not covered by the Pad\'e conjecture. In principle outliers are possible also for diagonal approximants, since only a subsequence of approximants is expected to converge. 

The efficiency of the Pad\'e resummation in recovering a robust and accurate approximation from the divergent perturbation series is impressive, in particular for $d=3$ and $4$, which are the extreme cases of monotonous divergence. We observe that Pad\'e approximants with at least $L+M \approx 10$ are needed to obtain a quantitative agreement with the NCSM result. In turn this shows that the information contained in the high-order energy corrections, which are responsible for the break-down of the power series, is indispensable to obtain stable and accurate results from the sequence of Pad\'e approximants.

\begin{figure}[t]
\includegraphics[width=1.05\columnwidth]{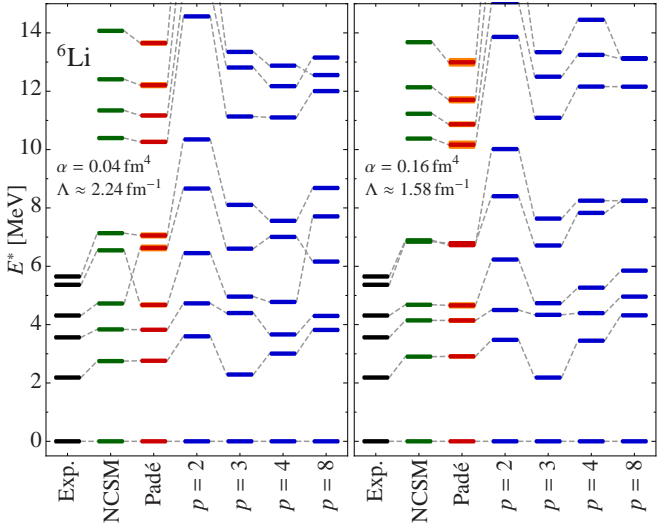}
\caption{(color online) Low-lying spectrum of \elem{Li}{6} computed in an $N_\text{max}=8$ model space for $\hbar\Omega=20\,\text{MeV}$, using SRG transformed N$^{3}$LO two-body interactions with flow parameter $0.04\,\text{fm}^{4} (\Lambda\approx2.24\,\text{fm}^{-1})$ in (a)  and  $0.16\,\text{fm}^{4} (\Lambda\approx1.58\,\text{fm}^{-1})$ in (b). Shown are from the left the  experimental observed energies, the NCSM results, the Pad\'e resummed results (see text for details) and results from the truncated power series of DMBPT at order $p=2$, 3, 4 and 8. Experimental values taken from Ref.~\cite{TiCh02}. }
\label{fig:spectrumLi6}
\end{figure}

In Fig.~\ref{fig:spectrumLi6} we show the excitation spectrum for the positive parity states of \elem{Li}{6} again for SRG parameters $\alpha=0.04\,\text{fm}^{4}$ (a) and $\alpha=0.16\,\text{fm}^{4}$ (b). The columns from left to right represent the experimental spectrum, the exact NCSM results, the results from the Pad\'e approximation and the spectrum obtained by truncating the power series at second, third, 4th, and 8th order. We extract an averaged Pad\'e result from the data shown in Fig.~\ref{fig:Pade_Li6} in the following way: As discussed earlier, the diagonal as well as the sub- and super-diagonal Pad\'e approximants $[L/M]$ for $L+M\gtrsim 15$ are very stable, except for extremely few outliers. Therefore, we compute the average of all approximants for $L+M\geq 15$ excluding those approximants which deviate by more than $0.5\,\text{MeV}$ from the average of the remaining set. We use the standard deviation of the approximants as a measure for the uncertainty of this average of the approximants. In Fig.~\ref{fig:spectrumLi6} the bars representing the Pad\'e results show an additional orange band representing this uncertainty. Overall, the uncertainty bands are very small and for the first 5 excited states we find remarkably good agreement between the Pad\'e resummed excitation energy and the exact results obtained in the NCSM. For the four high-lying states we observe a small deviation of the Pad\'e resummed result from the NCSM values, reaching about 500 keV for the highest state.

In contrast, the low-order DMBPT results do not provide a stable and reliable approximation for the exact eigenvalues. The excitation energies from second-order DMBPT are generally too large and the deviations from the exact NCSM results increase with increasing excitation energy. The third-order contribution typically lowers the excitation energy. In a few cases the exact eigenvalue appears between the second- and third-order estimate, however, in other cases both low-order estimates are still above the exact NCSM eigenvalue. The inclusion of the fourth-order  contribution typically improves the results, but going to still higher orders destroys the agreement again. All these statements hold also for the soft potential with $\alpha=0.16\,\text{fm}^4$, see Fig.~\ref{fig:spectrumLi6}(b).

\begin{figure}[t]
\includegraphics[width=1.05\columnwidth]{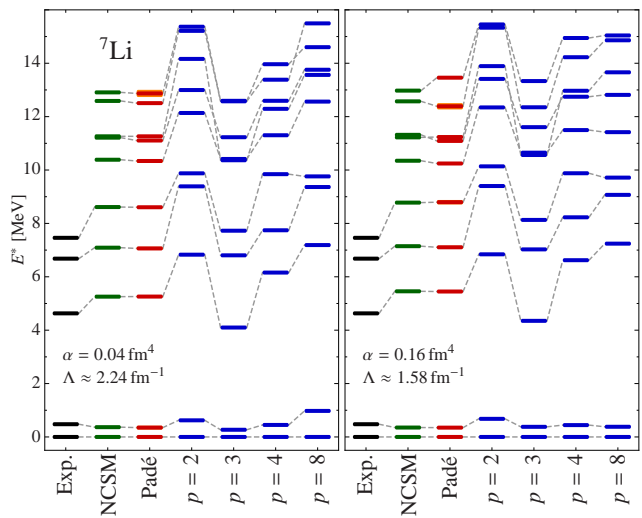}
\caption{(color online) Excitation energies of the nine energetically lowest states of the \elem{Li}{7} spectrum computed in an $N_\text{max}=8$ model space for $\hbar\Omega=20\,\text{MeV}$, using SRG transformed N$^{3}$LO two-body interactions with flow parameter $0.04\,\text{fm}^{4} (\Lambda\approx2.24\,\text{fm}^{-1})$ in (a) and  $0.16\,\text{fm}^{4} (\Lambda\approx1.58\,\text{fm}^{-1})$ in (b). Shown are from the left the  experimental observed energies, the NCSM results, the Pad\'e resumed results (see text for details) and results from the truncated power series of DMBPT at order $p=2$, 3, 4 and 8. Experimental values taken from Ref.~\cite{TiCh02}. }
\label{fig:spectrumLi7}
\end{figure}

A similar picture emerges for the negative-parity spectrum of \elem{Li}{7} depicted in Fig.~\ref{fig:spectrumLi7}, again using an $N_{\text{max}}=8$ model space for the NCSM and the DMBPT calculations. Again, we find excellent agreement of the Pad\'e-resummed energies with the exact NCSM spectrum for all states, with very stable Pad\'e approximations giving rise to very small uncertainties. The only exception is the highest excited state computed with the $\alpha=0.16\,\text{fm}^4$ interaction, that shows about 500\,keV deviation compared to the NCSM result. In contrast, the low-order DMBPT results show sizable deviations from the exact NCSM energies and change substantially from order to order. Again the second-order result typically overestimates the excitation energy and the third-order contribution lowers the excitation energy and often underestimates the excitation energy. With increasing order the changes become less coherent and for high orders, beyond the order $p=8$ the onset of divergence of the DMBPT series destroys the excitation spectrum completely (not shown in Figs.~\ref{fig:spectrumLi6} and \ref{fig:spectrumLi7}).

Finally, we note that though the general structure of the spectrum obtained in NCSM or Pad\'e-resummed DMBPT is in agreement with experiment for the low-lying states, the excitation energies are systematically to high. This hints at deficiencies of the SRG-evolved two-body interaction used here, e.g. the lack of a three-nucleon interaction. The inclusion of three-body forces is straight-forward, because the recursive formulas of section~\ref{sec:DMBPT} remain unchanged since the formalism is developed in terms of $A$-body Slater determinants. However, this is beyond the scope of this paper.

\section{Conclusions}

We have discussed degenerate Rayleigh-Schr\"odinger many-body perturbation theory for the description of ground and excited states of light nuclei with realistic Hamiltonians. We have derived an efficient, recursive formulation for the energy and state corrections in DMBPT that allows us to evaluate the perturbation series order-by-order up to very high orders, typically up to $p=30$. These formal developments pave the way for first direct applications of DMBPT to open-shell nuclei and excited states, where one has to deal with degeneracies by construction. 

We demonstrate the application of this formalism for the ground and excited states of \elem{Li}{6} and \elem{Li}{7} using an SRG-transformed NN interaction from chiral EFT at N$^3$LO. In order to allow for a direct comparison with NCSM calculations for the same model space, we use the harmonic oscillator as unperturbed Hamiltonian. We find that the perturbation series itself is divergent for all states considered although the interactions used here are extremely soft and are sometimes termed `perturbative'. This divergence of order-by-order MBPT was observed for ground states of closed-shell systems already in Ref. \cite{RoLa10}. To overcome this problem we resum the perturbation series using Pad\'e approximants, i.e., we use the energy corrections from DMBPT to construct a rational function instead of a simple power series to approximate the energy. We find that Pad\'e approximants using only low-order DMBPT results yield no improvement compared to low-order DMBPT results. However, if we include DMBPT information from high orders, i.e., 15th to 30th order, the diagonal and neighbor-diagonal Pad\'e approximants are in excellent agreement with exact NCSM calculations for the same model space. In contrast, low-order DMBPT results are clearly not sufficient to provide a quantitative and reliable approximation to the exact NCSM excitation energies.

A number of interesting topics remain, which we will address in the future: We can use the formalism of degenerate many-body perturbation theory to study binding-energy systematics also for heavier open-shell nuclei, eventually also including three-body forces. Moreover, the convergence pattern of the DMBPT series will be affected by the partitioning of the Hamiltonians, i.e. by the choice of the unperturbed basis. There are hints from preliminary studies in the nondegenerate case that, e.g., a Hartree-Fock basis can improve the convergence behavior of the corresponding power series.


\section*{Acknowledgments}

This work is supported by the Deutsche Forschungsgemeinschaft through contract SFB 634 and by the Helmholtz International Center for FAIR within the framework of the LOEWE program launched by the State of Hesse, and the BMBF through contract 06DA9040I.


\bibliography{/Users/joachim/uni/biblib/bib_nucl}

\end{document}